\documentclass[final,5p,times,twocolumn]{elsarticle}
\usepackage{lineno,hyperref,amsmath,amsthm,amssymb,amsfonts,ragged2e,color,subfig}
\journal{``Contributions to Plasma Physics"}
\begin{document}
\begin{frontmatter}
\title{Ion-acoustic rogue waves in double pair plasma having non-extensive particles}
\author{S. Jahan$^{*,1}$, M.N. Haque$^{**,1}$, N.A. Chowdhury$^{***,2}$, A. Mannan$^{\dag,3}$, and A.A. Mamun$^{\S,1}$}
\address{$^{1}$Department of Physics, Jahangirnagar University, Savar, Dhaka-1342, Bangladesh\\
$^2$ Plasma Physics Division, Atomic Energy Centre, Dhaka-1000, Bangladesh\\
$^3$ Institut f\"{u}r Mathematik, Martin Luther Universit\"{a}t Halle-Wittenberg, Halle, Germany\\
e-mail: $^*$jahan88phy@gmail.com, $^{**}$nhaque1219@gmail.com, $^{***}$nurealam1743phy@gmail.com,\\
$^{\dag}$abdulmannan@gmail.com, $^{\S}$mamun\_phys@juniv.edu}
\begin{abstract}
The modulational instability (MI) of ion-acoustic (IA) waves (IAWs) and associated IA rogue waves (IARWs) in double pair
plasma containing non-extensive electrons, iso-thermal positrons, negatively and positively charged
ions have been governed by the standard nonlinear Schr\"{o}dinger equation (NLSE). It has been figured
out from the numerical study of NLSE that the plasma system holds  modulationally
stable (unstable) region in which the dispersive and nonlinear coefficients of the NLSE have the
opposite (same) signs. It is also found that the fundamental features of IAWs (viz., MI criteria,
amplitude and width of the IARWs, etc.) are rigorously organized by the plasma parameters such as
mass, charge state, and number density of the plasma components. The existing outcomes of our present
study should be helpful for understanding the nonlinear features of IAWs (viz., MI and IARWs) 
in both laboratory and space plasmas.
\end{abstract}
\end{frontmatter}
\section{Introduction}
Double pair plasma (DPP) is characterised as fully ionized gas having  electrons, positron as well as
positive and negative ions, and is believed to exist in astrophysical environments such as
Van Allen radiation belt and near the polar cap of fast rotation neutron stars \cite{Lightman1982},
solar atmosphere \cite{Tandberg1988}, D-region ($H^+, O_2^-$) and F-region ($H^+, H^-$) of the
earths's ionosphere \cite{Elwakil2010}, upper region of Titan's atmosphere \cite{Labany2012} and
also in laboratory environments  \cite{Bacal1979,Gottoscho1986,Shukla1986,Berezhiani1992,Oohara2003,Helander2003}.
A number of authors \cite{Esfandyari-Kalejahi2006,Abdelsalam2008,Sabry2008} studied ion-acoustic (IA) waves (IAWs)
and associated nonlinear electrostatic structures namely, solitons, shocks, rogue waves, and double layers in the DPP.

Maxwellian distribution function is one of the most widely used velocity distribution functions of
particles to describe the dynamics of the iso-thermal particles. But it has been observed
that the characteristics of majority of particles in the space \cite{Young2005} and laboratory plasma
environments \cite{Lundin1989} are departed from the Maxwellian distribution. So, to narrate
the non-Maxwellian particles, Renyi \cite{Renyi1955} first recognized
the modification of Maxwellian distribution, and  finally, Tsallis \cite{Tsallis1988}
generalized the non-extensive $q$-distribution. It is noted that the index $q$ in the non-extensive $q$-distribution
characterizes the degree of non-extensivity of the particles \cite{Wang2013}.
Shalini \textit{et al.} \cite{Shalini2015} studied IAWs in non-extensive
plasma having two-temperature electrons, and observed that the width of the first
and second order IA rogue waves (IARWs) associated with IAWs decreases with increasing the
value of $q$ but the amplitude of the first and second-order IARWs associated with IAWs
is remain constant. Tribeche \textit{et al.} \cite{Tribeche2010} investigated electrostatic solitary waves
in presence of the non-extensive electrons, and found that the
amplitude of the potential increases with non-extensive parameter.
Hafez and Talukder \cite{Hafez2015} examined the propagation of the nonlinear electrostatic waves in a
three-component non-extensive plasma having inertialess non-extensive electrons and positrons, and inertial
ions, and reported that the amplitude of the soliton increases with increasing temperature of the
non-extensive electron.

The investigation of the modulational instability (MI) \cite{C1,C2,C3,C4,C5,C6} and associated nonlinear features of
wave is one of the most important research areas for plasma physicists. It is noteworthy that
the MI of the wave is considered to be the primary reason for the formation of massive and
gigantic rogue waves (RWs) \cite{Akhmediev2009}. Rogue wave, which is the rational solution
of the standard nonlinear Schr\"{o}dinger equation (NLSE) \cite{Akhmediev2009,Kedziora2011,Haque2019,Haque2019a},
is a short-lived phenomenon which emerges from nowhere and disappears without a trace \cite{Kedziora2011}.
A number of authors have investigated the MI of IAWs by considering the non-extensive particles \cite{Bains2011,Bouzit2015,Eslami2011}.
Bains \textit{et al.} \cite{Bains2011} studied the MI of IAWs in presence of non-extensive electrons,
and demonstrated that the critical wave number ($k_c$) at which the instability sets in increases with the increase in the value of $q$ ($q>0$).
Bouzit \textit{et al.} \cite{Bouzit2015} investigated the stability conditions of IAWs
in presence of non-extensive non-thermal electrons. Eslami \textit{et al.} \cite{Eslami2011}
investigated the MI of IAWs in electron-positron-ion plasma having non-extensive electrons and positrons,
and observed that the $k_c$ decreases with $q$ for $q<0$ while increases with $q$ for $q>0$.
To the best knowledge of the authors, no attempt has been made to investigate the MI of the IAWs
and associated IARWs in a four-component plasma containing inertial positively and negatively charged ions,
and inertialess non-extensive electrons, and iso-thermal positrons. Therefore, it is a rational
fascination to examine the influence of non-extensive electrons and iso-thermal positrons on
the MI of IAWs and associated IARWs in a four-component DPP.

The manuscript is organized as the following pattern: The model equations are presented
in Sec. \ref{3sec:Model Equations}. The derivation of the NLSE is shown in Sec. \ref{3sec:Derivation of the NLSE}.
The stability of IAWs is provided in Sec. \ref{3sec:Stability of IAWs}. The IARWs is demonstrated  in Sec. \ref{3sec:Rogue Waves}.
Finally, a conclusion is given in Sec. \ref{3sec:Conclusion}.
\section{Model Equations}
\label{3sec:Model Equations}
We consider the propagation of IAWs in a collisionless, fully ionized, unmagnetized
plasma system consisting of warm negative ions, symbolized by $n_{-i}$
(charge $q_{-i}=-Z_{-i}e$; mass $m_{-i}$), warm positive ions, denoted by $n_{+i}$ (charge $q_{+i}=Z_{+i}e$; mass $m_{+i}$),
non-extensive $q$-distributed electrons, identified by $n_e$ (charge $q_e=-e$; mass $m_e$), and  iso-thermal positrons, expressed
by $n_p$ (charge $q_p=+e$; mass $m_p$); where $Z_{-i}$ ($Z_{+i}$) is the charge state of the negatively (positively) charged ion, and $e$ being the magnitude of the charge of the electron.
The charge neutrality condition of our present model can be written as
$n_{p0}+Z_{+i} n_{+i0}=n_{e0}+Z_{-i} n_{-i0}$.
Now, the  normalized equations can be given in the following form
\begin{eqnarray}
&&\hspace*{-1.3cm}\frac{\partial n_{-i}}{\partial t}+\frac{\partial}{\partial x}(n_{-i} u_{-i})=0,
\label{3eq:1}\\
&&\hspace*{-1.3cm}\frac{\partial u_{-i}}{\partial t} + u_{-i}\frac{\partial u_{-i}}{\partial x}+\lambda_1 n_{-i} \frac{\partial n_{-i}}{\partial x}= \frac{\partial \phi}{\partial x},
\label{3eq:2}\\
&&\hspace*{-1.3cm}\frac{\partial n_{+i}}{\partial t}+\frac{\partial}{\partial x}(n_{+i} u_{+i})=0,
\label{3eq:3}\\
&&\hspace*{-1.3cm}\frac{\partial u_{+i}}{\partial t} + u_{+i}\frac{\partial u_{+i}}{\partial x}+\lambda_3 n_{+i} \frac{\partial n_{+i}}{\partial x}=-\lambda_2 \frac{\partial \phi}{\partial x},
\label{3eq:4}\\
&&\hspace*{-1.3cm}\frac{\partial^2\phi}{\partial x^2}=\lambda_4 n_e+n_{-i}-(1+\lambda_4-\lambda_5)n_p-\lambda_5n_{+i},
\label{3eq:5}\
\end{eqnarray}
where $n_{-i}$, $n_{+i}$, $n_e$, and $n_p$ are normalized by $n_{-i0}$, $n_{+i0}$, $n_{e0}$, and $n_{p0}$,
respectively; $u_{-i}$ and $u_{+i}$ indicate the negatively and positively charged ion fluid, respectively,
normalized by the IA wave speed $C_{-i}=(Z_{-i}k_BT_e/m_{-i})^{1/2}$ (with $k_B$ being the Boltzmann constant
and $T_e$ being the temperature of the electron); $\phi$ denoted as the electrostatic wave potential,
normalized by $k_BT_e/e$; the time and space variables are, respectively normalized by $\omega_{-ip}^{-1}=(m_{-i}/4 \pi e^2Z_{-i}^2 n_{-i0})^{1/2}$ and
$\lambda_{-i}=(k_BT_e/4 \pi e^2 Z_{-i} n_{-i0})^{1/2}$. The pressure term of the ion can be represented as $P_{-i}=P_{-i0}(N_{-i}/n_{-i0})^\gamma$ with $P_{-i0}=n_{-i0}k_BT_{-i}$; where $P_{-i0}$ ($T_{-i}$) being the equilibrium pressure (temperature) of the negatively charged ion, and $P_{+i}=P_{+i0}(N_{+i}/n_{+i0})^\gamma$ with $P_{+i0}=n_{+i0}k_BT_{+i}$; where $P_{+i0}$ ($T_{+i}$) being the equilibrium pressure (temperature) of the positively charged ion, respectively, and $\gamma=(N+2)/N$ (where $N$ recognized as the degree of freedom and for one-dimensional case $N=1$, so $\gamma=3$). Other parameters can be defined as $\lambda_1=3T_{-i}/Z_{-i}T_e$, $\lambda_2=Z_{+i}m_{-i}/Z_{-i}m_{+i}$, $\lambda_3=3T_{+i}m_{-i}/Z_{-i}m_{+i}T_e$, $\lambda_4=n_{e0}/Z_{-i}n_{-i0}$, and $\lambda_5=Z_{+i}n_{+i0}/Z_{-i}n_{-i0}$.
Now, the number densities of the non-extensive $q-$distributed \cite{Tsallis1988,Jahan2019}
electron and iso-thermally distributed \cite{Shukla2002,Chowdhury2019a} positron can be represented by the following normalized equations
\begin{eqnarray}
&&\hspace*{-1.3cm}n_e=[1+(q-1)\phi]^{\frac{q+1}{2(q-1)}},
\label{3eq:6}\\
&&\hspace*{-1.3cm}n_p= \mbox{exp}(-\lambda_6 \phi),
\label{3eq:7}\
\end{eqnarray}
where  $\lambda_6=T_e/T_p$ (with $T_p$ being the temperature of the iso-thermally distributed
positron and $T_e>T_p$). The parameter $q$, generally known as entropic index which quantifies the degree of non-extensivity. It is noteworthy that when $q=1$, the entropy reduces to standard Maxwell-Boltzmann distribution. On the other hand, in the limits $q>0$ ($q<0$), the entropy shows sub-extensivity (super-extensivity). Now, by substituting Eqs. $(6)$ and $(7)$ into Eq. $(5)$ and expanding up to third order in $\phi$, we can draw up as
\begin{eqnarray}
&&\hspace*{-1.3cm}\frac{\partial^2\phi}{\partial x^2}+1+\lambda_5 n_{+i}=\lambda_5+n_{-i}+M_1\phi+M_2\phi^2+M_3\phi^3+\cdot\cdot\cdot,
\label{3eq:8}\
\end{eqnarray}
where
\begin{eqnarray}
&&\hspace*{-1.3cm}M_1=[\lambda_4(q+1)+2\lambda_6(1+\lambda_4-\lambda_5)]/2,
\nonumber\\
&&\hspace*{-1.3cm}M_2=[\lambda_4(q+1)(3-q)-4\lambda_6^2(1+\lambda_4-\lambda_5)]/8,
\nonumber\\
&&\hspace*{-1.3cm}M_3=[\lambda_4(q+1)(q-3)(3q-5)+8\lambda_6^3(1+\lambda_4-\lambda_5)]/48.
\nonumber\
\end{eqnarray}
It is noted that the terms containing $M_1$, $M_2$, and $M_3$  in Eq. \eqref{3eq:8} are due to
the contribution of the non-extensive $q$-distributed electrons and iso-thermal positrons.
\section{Derivation of the NLSE}
\label{3sec:Derivation of the NLSE}
To study  the MI of the IAWs, first we want to construct the NLSE by employing the reductive perturbation method.
In that case, the stretched co-ordinates can be written in the following fashion \cite{Jahan2019,Chowdhury2017a,C7,C8,C9,C10}
\begin{eqnarray}
&&\hspace*{-1.3cm}\xi=\epsilon(x-v_g t),
\label{3eq:9}\\
&&\hspace*{-1.3cm}\tau=\epsilon^2t,
\label{3eq:10}\
\end{eqnarray}
where $v_g$ is the group speed and $\epsilon$ is a small parameter. After that the dependent variables 
can be represented as \cite{Jahan2019,Chowdhury2017a,C7,C8,C9,C10}
\begin{eqnarray}
&&\hspace*{-1.3cm}\Pi(x,t)=\Pi_0+\sum_{m=1}^\infty \epsilon^{(m)} \sum_{l=-\infty}^\infty \Pi_l^{(m)}(\xi,\tau)M,
\label{3eq:11}\
\end{eqnarray}
where $\Pi_{il}^{(m)}=[n_{-il}^m, u_{-il}^m, n_{+il}^m, u_{+il}^m, \phi_{il}^m]$, $\Pi_0=[1,0,1,0,0]^T$, $M=\mbox{exp}[il(kx-\omega t)]$, and $k~(\omega)$ is real variables representing the carrier wave number (frequency). The derivative operators can be showed as
\begin{eqnarray}
&&\hspace*{-1.3cm}\frac{\partial}{\partial t}\rightarrow\frac{\partial}{\partial t}-
\epsilon v_g \frac{\partial}{\partial\xi}+\epsilon^2\frac{\partial}{\partial\tau},
\label{3eq:12}\\
&&\hspace*{-1.3cm}\frac{\partial}{\partial x}\rightarrow\frac{\partial}{\partial x}+
\epsilon\frac{\partial}{\partial\xi}.
\label{3eq:13}\
\end{eqnarray}
Now, by substituting Eqs. \eqref{3eq:9}$-$\eqref{3eq:13} into Eqs. \eqref{3eq:1}$-$\eqref{3eq:4},
and \eqref{3eq:8}, and collecting power term of $\epsilon$, the first order ($m=1$ with $l=1$) shortened equations can be presented as
\begin{eqnarray}
&&\hspace*{-1.3cm}n_{-i1}^{(1)}=\frac{k^2}{S}\phi_1^{(1)},
\label{3eq:14}\\
&&\hspace*{-1.3cm}u_{-i1}^{(1)}=\frac{k\omega}{S}\phi_1^{(1)},
\label{3eq:15}\\
&&\hspace*{-1.3cm}n_{+i1}^{(1)}=\frac{\lambda_2k^2}{A}\phi_1^{(1)},
\label{3eq:16}\\
&&\hspace*{-1.3cm}u_{+i1}^{(1)}=\frac{\lambda_2 \omega k}{A}\phi_1^{(1)},
\label{3eq:17}\
\end{eqnarray}
where $S=\lambda_1k^2-\omega^2$ and $A=\omega^2-\lambda_3k^2$.
These equations provide the dispersion relation of IAWs in the following form
\begin{eqnarray}
&&\hspace*{-1.3cm}\omega^2=\frac{I \pm \sqrt{I^2-4UJ}}{2U},
\label{3eq:18}
\end{eqnarray}
where $I=(\lambda_1 k^2+\lambda_3 k^2+\lambda_1 M_1+\lambda_3 M_1+\lambda_2 \lambda_5+1)$, $U=(k^2+M_1)/k^2$,
and $J=k^2(\lambda_1 \lambda_3 k^2+\lambda_1 \lambda_3 M_1+\lambda_3+\lambda_1 \lambda_2 \lambda_5)$. However,
to obtain the real and positive values of $\omega$, the conditions $I^2>4UJ$ must be maintained. It is noted
that the positive and negative signs in Eq. \eqref{3eq:18} resembled to the fast ($\omega_f$) and slow ($\omega_s$)
IA modes, respectively. The second order ($m=2$ with $l=1$) equations and with the compatibility condition, we can be written the
group speed of the IAWs
\begin{eqnarray}
&&\hspace*{-1.3cm}v_g=\frac{(L+\lambda_1A^2k^2+A^2\omega^2-2A^2S^2-SA^2)}{2\omega k(A^2+\lambda_2\lambda_5S^2)},
\label{3eq:23}\
\end{eqnarray}
where $L=\lambda_2\lambda_3\lambda_5k^2S^2+\lambda_2\lambda_5\omega^2S^2+\lambda_2\lambda_5AS^2$.
Now, the co-efficient of $\epsilon$ (when $m=2$ with $l=2$) yield the second-order harmonic amplitudes
are found to be proportional to $|\phi_1^{(1)}|^2$
\begin{eqnarray}
&&\hspace*{-1.3cm}n_{-i2}^{(2)}=M_4|\phi_1^{(1)}|^2,
\label{3eq:24}\\
&&\hspace*{-1.3cm}u_{-i2}^{(2)}=M_5|\phi_1^{(1)}|^2,
\label{3eq:25}\\
&&\hspace*{-1.3cm}n_{+i2}^{(2)}=M_6|\phi_1^{(1)}|^2,
\label{3eq:26}\\
&&\hspace*{-1.3cm}u_{+i2}^{(2)}=M_7|\phi_1^{(1)}|^2,
\label{3eq:27}\\
&&\hspace*{-1.3cm}\phi_{2}^{(2)}=M_8|\phi_1^{(1)}|^2,
\label{3eq:28}\
\end{eqnarray}
where
\begin{eqnarray}
&&\hspace*{-1.3cm}M_4=\frac{2 M_8 k^2 S^2-3\omega^2 k^4-\lambda_1 k^6}{2 S^3},
\nonumber\\
&&\hspace*{-1.3cm}M_5=\frac{\omega M_4 S^2-\omega k^4}{k S^2},
\nonumber\\
&&\hspace*{-1.3cm}M_6=\frac{2\lambda_2 M_8 A^2 k^2+3\lambda_2^2\omega^2 k^4+\lambda_3\lambda_2^2 k^6}{2 A^3},
\nonumber\\
&&\hspace*{-1.3cm}M_7=\frac{\omega M_6 A^2-\omega\lambda_2^2 k^4}{k A^2},
\nonumber\\
&&\hspace*{-1.3cm}M_8=\frac{2 M_2 A^3 S^3-3\lambda_5\lambda_2^2\omega^2 S^3 k^4-M_9}{M_{10}},
\nonumber\\
&&\hspace*{-1.3cm}M_9=3\omega^2 A^3 k^4+\lambda_3\lambda_5\lambda_2^2 S^3 k^6+\lambda_1 A^3 k^6,
\nonumber\\
&&\hspace*{-1.3cm}M_{10}=2 A^2 S^2(S\lambda_2\lambda_5k^2-4 k^2 A S-M_1 A S-A k^2).
\nonumber\
\end{eqnarray}
Next, consider the image for $m=3$ with $l=0$ and $m=2$ with $l=0$, which margined the zeroth harmonic modes.
In such a way we can get the following results
\begin{eqnarray}
&&\hspace*{-1.3cm}n_{-i0}^{(2)}=M_{11}|\phi_1^{(1)}|^2,
\label{3eq:29}\\
&&\hspace*{-1.3cm}u_{-i0}^{(2)}=M_{12}|\phi_1^{(1)}|^2,
\label{3eq:30}\\
&&\hspace*{-1.3cm}n_{+i0}^{(2)}=M_{13}|\phi_1^{(1)}|^2,
\label{3eq:31}\\
&&\hspace*{-1.3cm}u_{+i0}^{(2)}=M_{14}|\phi_1^{(1)}|^2,
\label{3eq:32}\\
&&\hspace*{-1.3cm}\phi_{0}^{(2)}=M_{15}|\phi_1^{(1)}|^2,
\label{3eq:33}\
\end{eqnarray}
where
\begin{eqnarray}
&&\hspace*{-1.3cm}M_{11}=\frac{2\omega v_g k^3+\lambda_1k^4+k^2\omega^2-M_{15}S^2}{S^2(v_g^2-\lambda_1)},
\nonumber\\
&&\hspace*{-1.3cm}M_{12}=\frac{M_{11} v_g S^2-2 \omega k^3}{S^2},
\nonumber\\
&&\hspace*{-1.3cm}M_{13}=\frac{\lambda_2M_{15}A^2+\lambda_3\lambda_2^2 k^4+\lambda_2^2 \omega^2 k^2+2\omega v_g\lambda_2^2 k^3}{A^2(v_g^2-\lambda_3)},
\nonumber\\
&&\hspace*{-1.3cm}M_{14}=\frac{M_{13} v_g A^2-2 \omega \lambda_2^2 k^3}{A^2},
\nonumber\\
&&\hspace*{-1.3cm}M_{15}=\frac{A^2(v_g^2-\lambda_3)(2\omega v_g k^3+\lambda_1 k^4+\omega^2 k^2)+M_{16}}{A^2 S^2v_g^2+M_{17}},
\nonumber\\
&&\hspace*{-1.3cm}M_{16}=2 M_2 A^2S^2(v_g^2-\lambda_1)(v_g^2-\lambda_3)-(v_g^2-\lambda_1)
\nonumber\\
&&\hspace*{-0.4cm}S^2(2 \lambda_5 \omega v_g\lambda_2^2 k^3+\lambda_3\lambda_5\lambda_2^2 k^4+\lambda_5\lambda_2^2\omega^2 k^2),
\nonumber\\
&&\hspace*{-1.3cm}M_{17}=A^2 S^2[\lambda_2\lambda_5(v_g^2-\lambda_1)-\lambda_3-M_1(v_g^2-\lambda_1)(v_g^2-\lambda_3)].
\nonumber\
\end{eqnarray}
Lastly, the third-order harmonic modes ($m=3$ with $l=1$), with the assistance of
Eqs. \eqref{3eq:14}$-$\eqref{3eq:33}, represent a complete set of equations, which
can be transformed to the NLSE:
\begin{eqnarray}
&&\hspace*{-1.3cm}i\frac{\partial \Phi}{\partial\tau}+P\frac{\partial^2\Phi}{\partial\xi^2}+Q|\Phi|^2\Phi=0,
\label{3eq:34}\
\end{eqnarray}
where $\Phi=\phi_1^{(1)}$ for simplicity. In Eq. \eqref{3eq:34}, $P$ is the dispersion co-efficient, which can be written as
\begin{eqnarray}
&&\hspace*{-1.3cm}P=\frac{4 \lambda_2 \lambda_5 \omega^2 k^2 v_g^2 S^3+4 \lambda_1 \omega v_g A^3 k^3+4kv_g \omega^3 A^3+M_{18}}{2 \omega A S k^2 (A^2+\lambda_2 \lambda_5 S^2)},
\nonumber\
\end{eqnarray}
\begin{eqnarray}
&&\hspace*{-1.3cm}M_{18}=2 \lambda_2 \lambda_3 \lambda_5 \omega^2 k^2 S^3+\lambda_2 \lambda_5 \lambda_3^2 S^3 k^4+\lambda_1 S k^2 A^3
\nonumber\\
&&\hspace*{-0.5cm}+\lambda_2\lambda_5 S^3\omega^4+\lambda_2\lambda_3\lambda_5 A k^2 S^3-4\lambda_2\lambda_5kv_g\omega^3 S^3
\nonumber\\
&&\hspace*{-0.5cm}-4 \lambda_2\lambda_3\lambda_5\omega v_g k^3 S^3-4\omega^2 k^2 v_g^2 A^3-\lambda_1^2 A^3 k^4
\nonumber\\
&&\hspace*{-0.5cm}-2 \lambda_1\omega^2 k^2 A^3-S k^2 v_g^2 A^3-A^3\omega^4
\nonumber\\
&&\hspace*{-0.5cm}-\lambda_2 \lambda_5 A k^2 v_g^2 S^3-A^3 S^3,
\nonumber\
\end{eqnarray}
and $Q$ is the nonlinear co-efficient, which can be written as
\begin{eqnarray}
&&\hspace*{-1.3cm}Q=\frac{3M_3 A^2 S^2+2 M_2 M_8 A^2 S^2+2 M_2 M_{15} A^2 S^2-M_{19}}{2 \omega k^2 (A^2+\lambda_2 \lambda_5 S^2)},
\nonumber\
\end{eqnarray}
where
\begin{eqnarray}
&&\hspace*{-1.3cm}M_{19}=2\omega M_5 A^2 k^3+2\lambda_2 \lambda_5\omega M_7 S^2 k^3+2\omega M_{12} A^2 k^3
\nonumber\\
&&\hspace*{-0.5cm}+2 \lambda_2 \lambda_5 \omega M_{14} S^2 k^3+\lambda_1 M_4 A^2 k^4+M_4\omega^2 A^2 k^2
\nonumber\\
&&\hspace*{-0.5cm}+\lambda_2 \lambda_3 \lambda_5 M_6 S^2 k^4+\lambda_2 \lambda_5 M_6 \omega^2 k^2 S^2+M_{11}\omega^2 A^2 k^2
\nonumber\\
&&\hspace*{-0.5cm}+\lambda_1 M_{11} A^2 k^4+\lambda_2 \lambda_5 M_{13}\omega^2 k^2 S^2+\lambda_2 \lambda_3 \lambda_5 M_{13}S^2 k^4
\nonumber
\end{eqnarray}
It may be noted here that both $P$ and $Q$ are directly depend on different parameters namely $\lambda_1$,
$\lambda_2$, $\lambda_3$, $\lambda_4$, $\lambda_5$, $\lambda_6$, $q$, and are indirectly depend on
mass, number density, temperature, and charge state of the different plasma components.
\begin{figure}[h!]
\centering
\includegraphics[width=75mm]{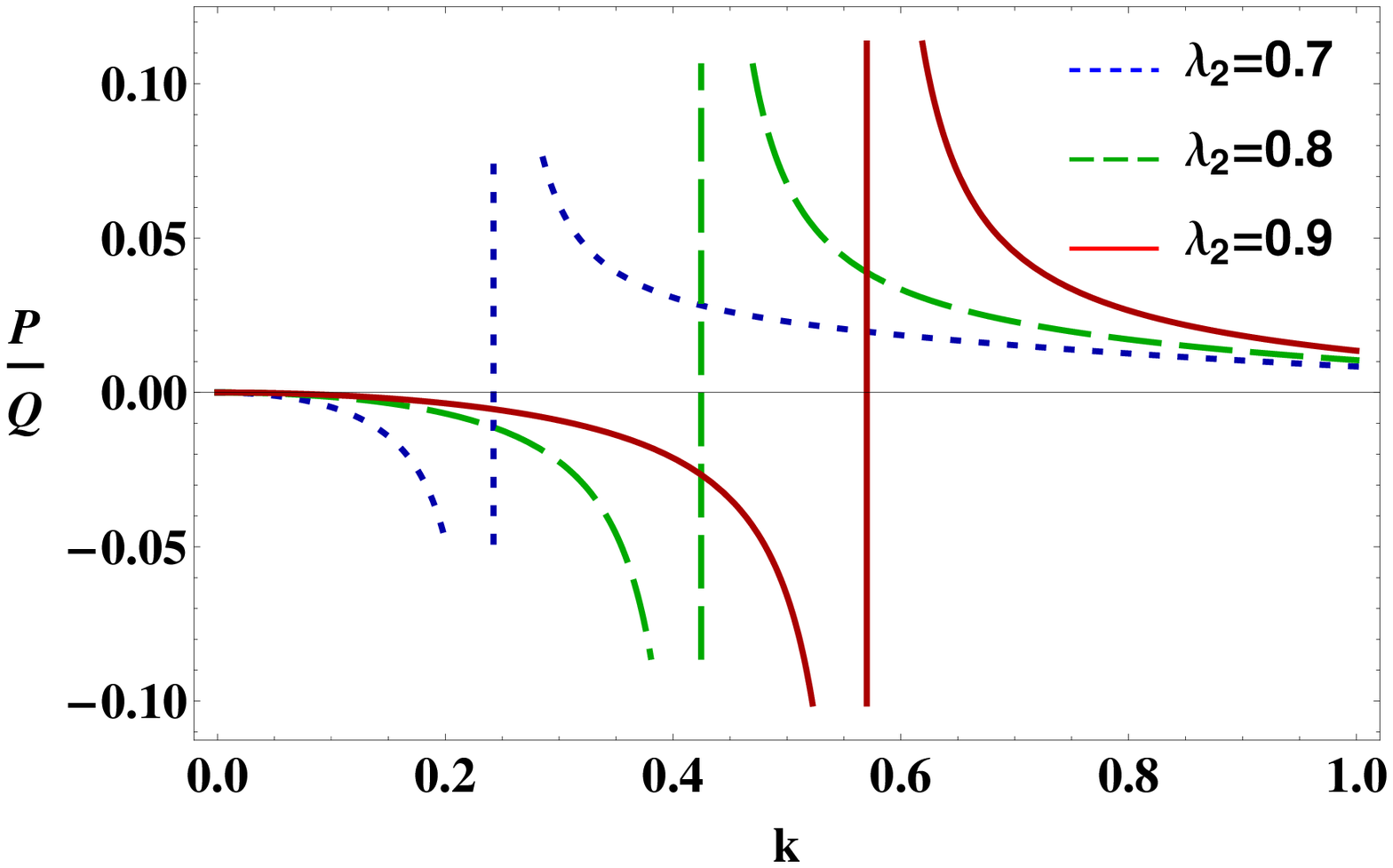}
\caption{The variation of $P/Q$ with $k$ for different values of $\lambda_2$
when $\lambda_1=0.007$, $\lambda_3=0.07$, $\lambda_4=1.8$, $\lambda_5=1.5$,
$\lambda_6=1.7$, $q=1.4$, and $\omega\equiv\omega_f$.}
\label{3Fig:F1}
\end{figure}
\begin{figure}[h!]
\centering
\includegraphics[width=75mm]{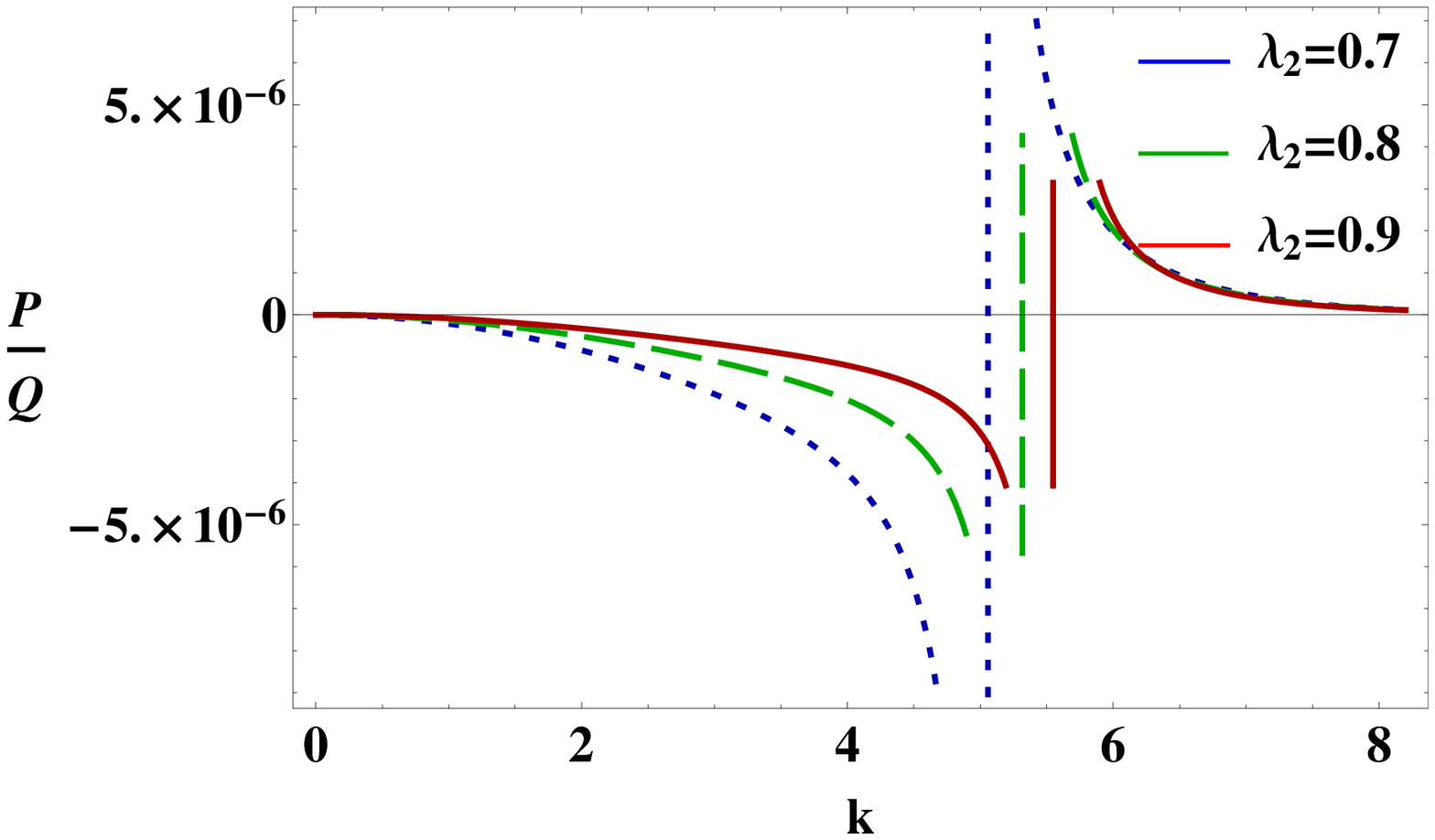}
\caption{The variation of $P/Q$ with $k$ for different values of $\lambda_2$
when $\lambda_1=0.007$, $\lambda_3=0.07$, $\lambda_4=1.8$, $\lambda_5=1.5$,
$\lambda_6=1.7$, $q=1.4$, and $\omega\equiv\omega_s$.}
\label{3Fig:F2}
\end{figure}
\begin{figure}[h!]
\centering
\includegraphics[width=80mm]{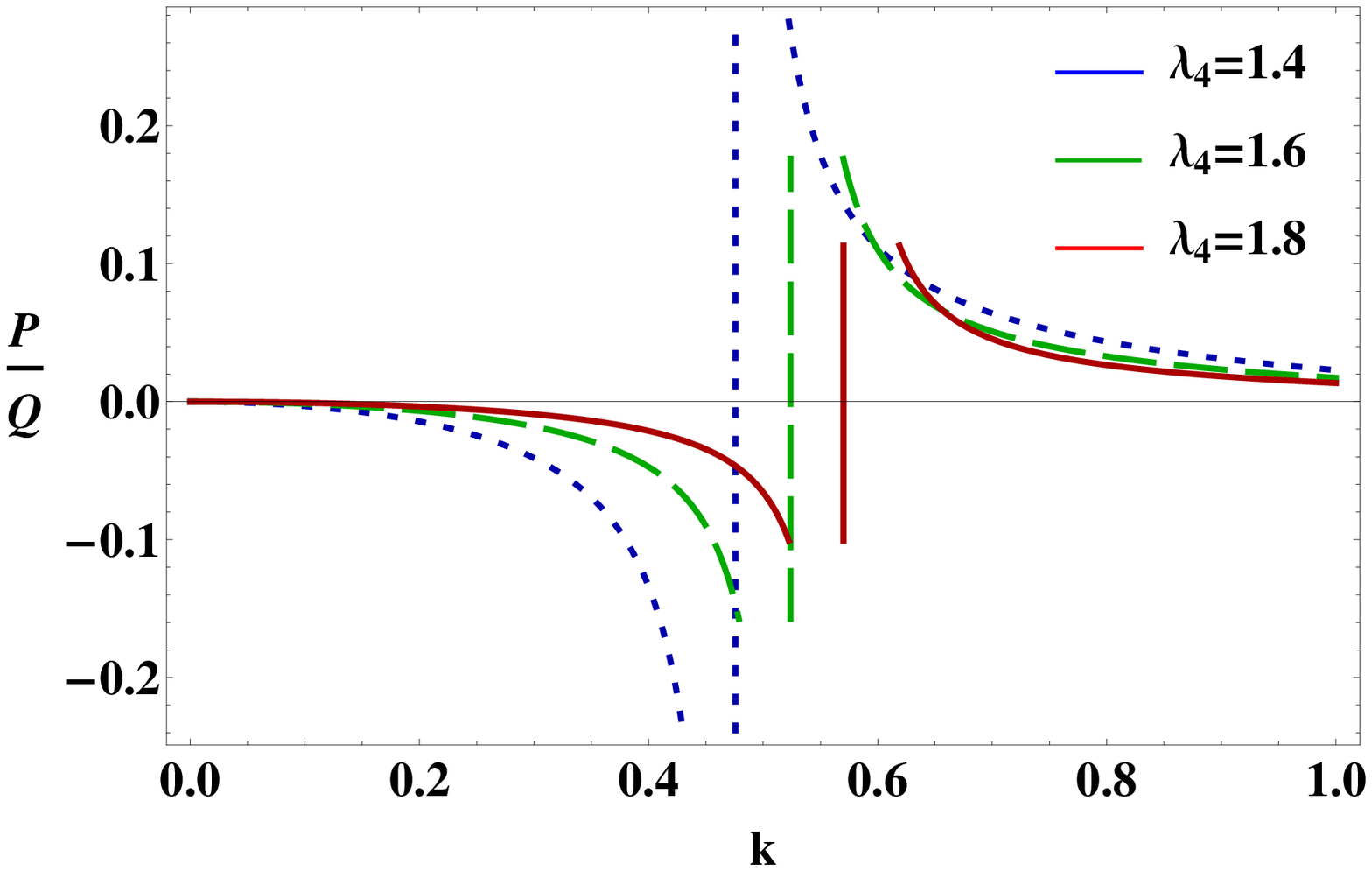}
\caption{The variation of $P/Q$ with $k$ for different values of $\lambda_4$
when $\lambda_1=0.007$, $\lambda_2=1.2$, $\lambda_3=0.07$, $\lambda_5=1.5$,
$\lambda_6=1.7$, $q=1.4$, and $\omega\equiv\omega_f$.}
\label{3Fig:F3}
\end{figure}
\begin{figure}[h!]
\centering
\includegraphics[width=80mm]{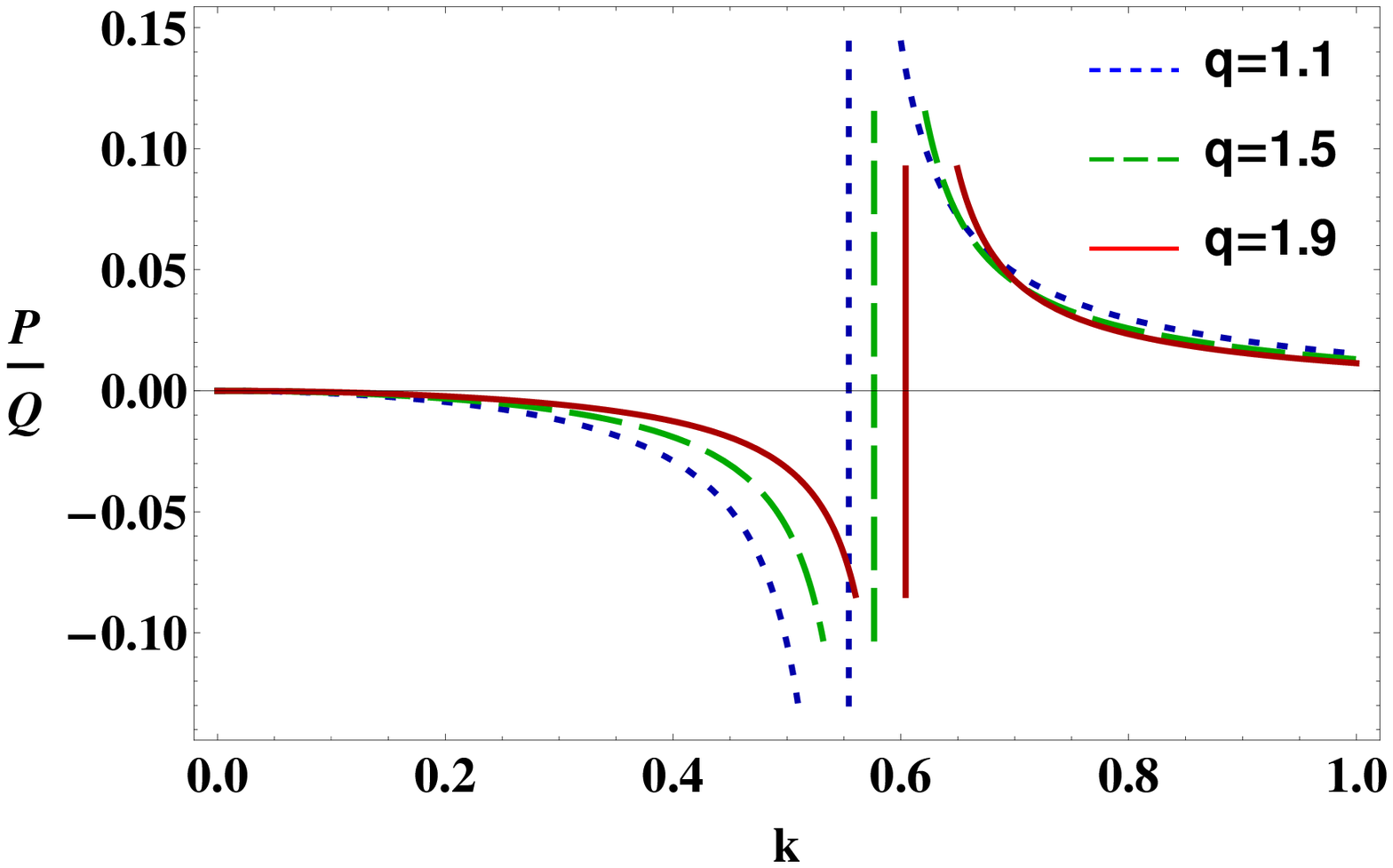}
\caption{The variation of $P/Q$ with $k$ for different values of $q$ when
$\lambda_1=0.007$, $\lambda_2=1.2$, $\lambda_3=0.07$, $\lambda_4=1.8$, $\lambda_5=1.5$,
$\lambda_6=1.7$, and $\omega\equiv\omega_f$.}
\label{3Fig:F4}
\end{figure}
\begin{figure}[h!]
\centering
\includegraphics[width=80mm]{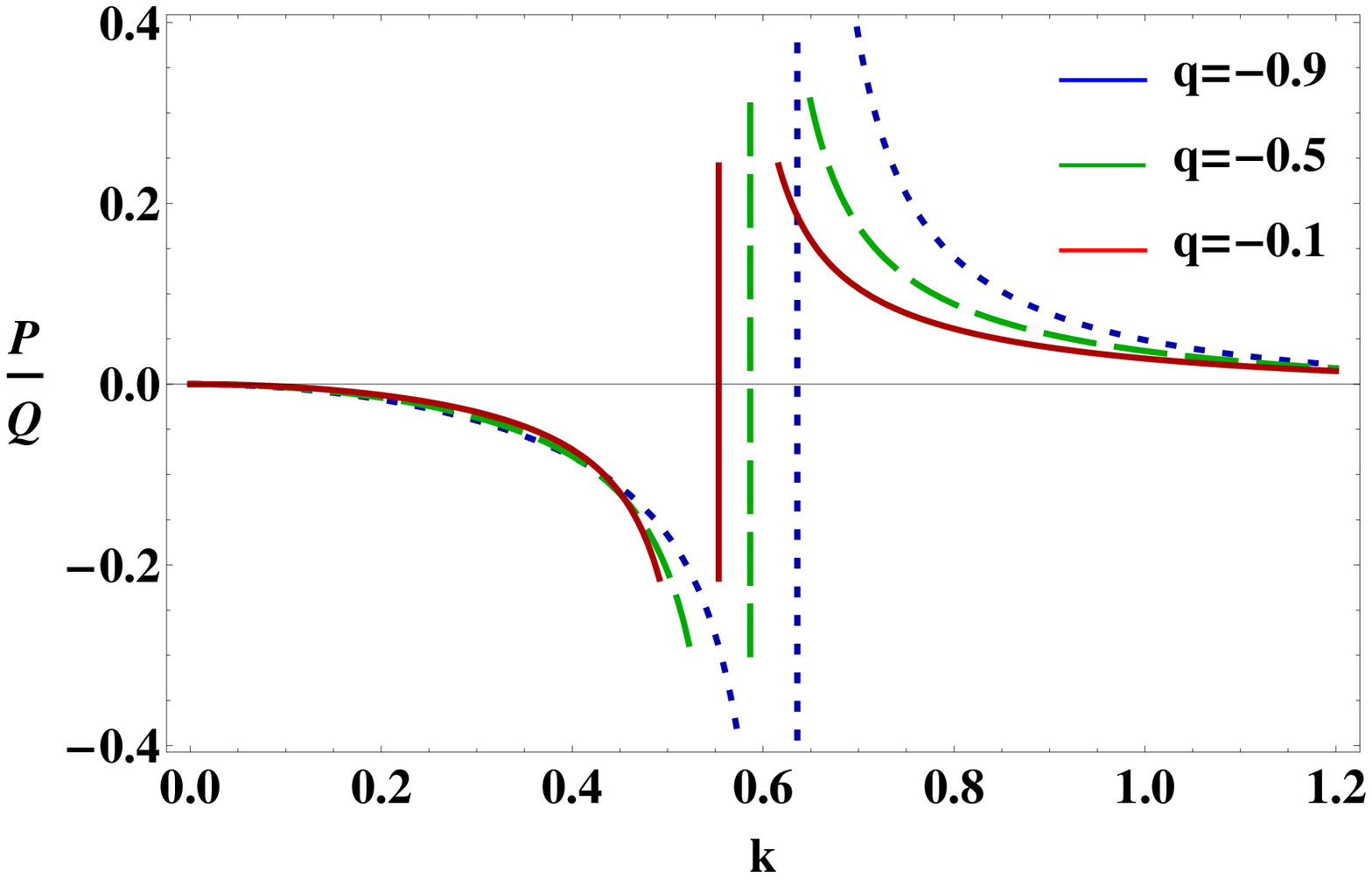}
\caption{The variation of $P/Q$ with $k$ for different values of $q$ when
$\lambda_1=0.007$, $\lambda_2=1.2$, $\lambda_3=0.07$, $\lambda_4=1.8$,
$\lambda_5=1.5$, $\lambda_6=1.7$, and $\omega\equiv\omega_f$.}
\label{3Fig:F5}
\end{figure}
\begin{figure}[h!]
\centering
\includegraphics[width=80mm]{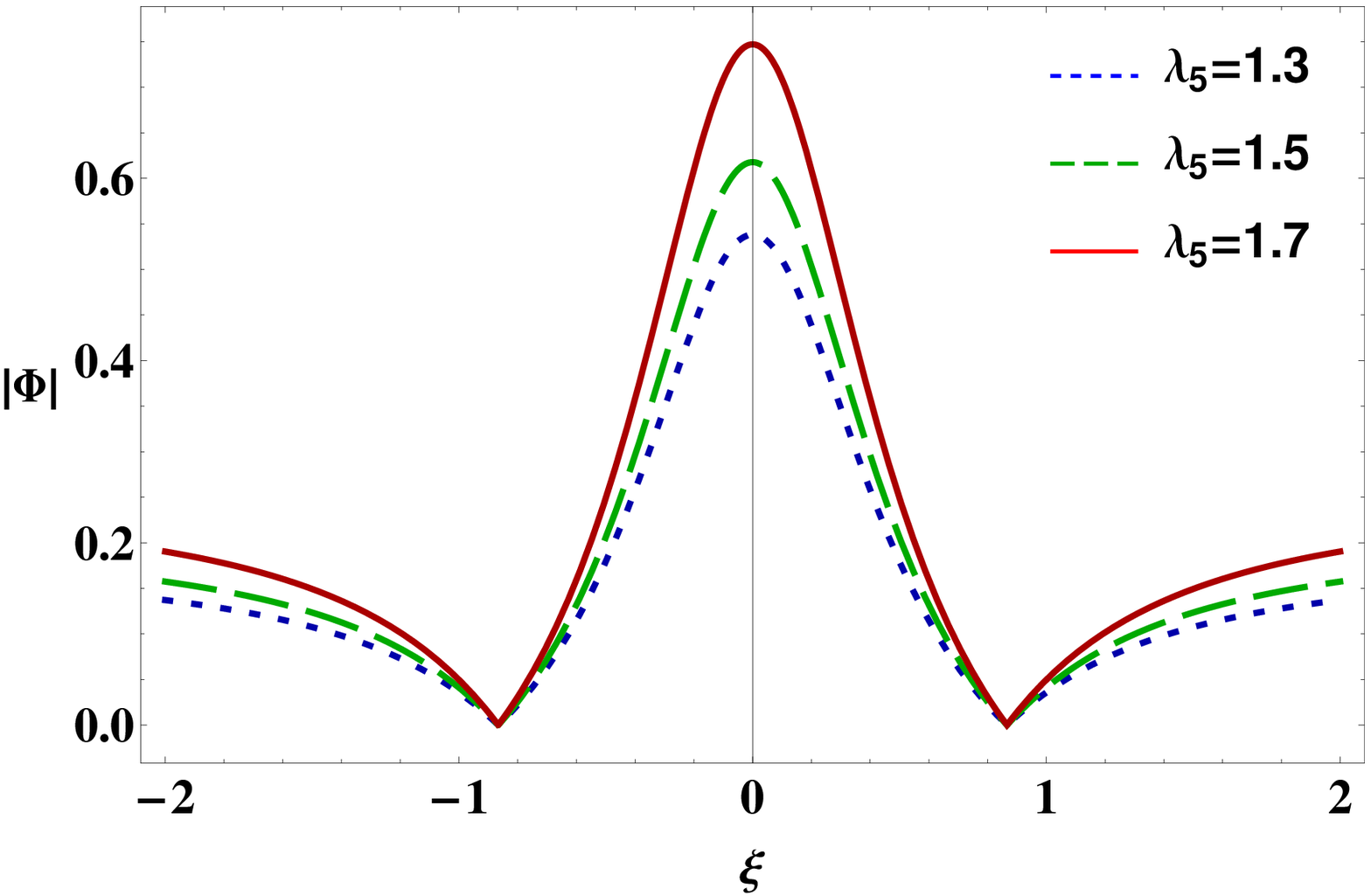}
\caption{The variation of $|\Phi|$ with $\xi$ for different values of $\lambda_5$
when $\lambda_1=0.007$, $\lambda_2=1.2$, $\lambda_3=0.07$, $\lambda_4=1.8$, $\lambda_6=1.7$,
$\tau=0$, $k=0.4$, and $\omega\equiv\omega_f$.}
\label{3Fig:F6}
\end{figure}
\begin{figure}[h!]
\centering
\includegraphics[width=80mm]{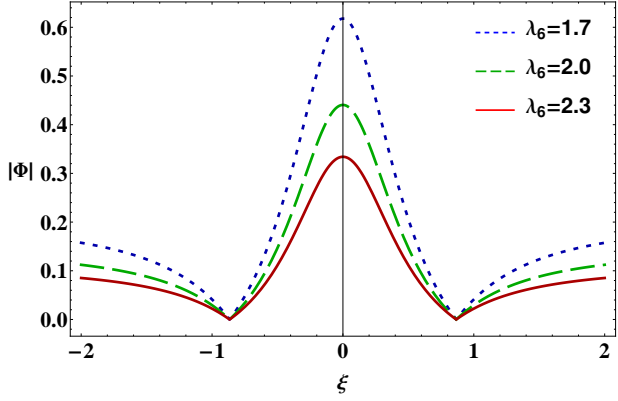}
\caption{The variation of $|\Phi|$ with $\xi$ for different values of $\lambda_6$  when
$\lambda_1=0.007$, $\lambda_2=1.2$, $\lambda_3=0.07$, $\lambda_4=1.8$, $\lambda_5=1.5$,
$\tau=0$, $k=0.4$, and $\omega\equiv\omega_f$.}
\label{3Fig:F7}
\end{figure}
\section{Stability of IAWs}
\label{3sec:Stability of IAWs}
The propagation of IAWs is modulationaly stable when $P$ and $Q$ have opposite sign (i.e., $P/Q<0$), and is
modulationally unstable when both $P$ and $Q$ have same sign (i.e., $P/Q>0$). The point at which the transition
of $P/Q$ curve intersect with the $k$-axis in the ``$P/Q$ versus $k$'' graph is known as threshold or critical
wave number $k~(=k_c)$.  Under consideration of fast and slow IA modes, we have depicted the $P/Q$ versus $k$
curve for different values of $\lambda_2$ in Figs. \ref{3Fig:F1} and \ref{3Fig:F2}, respectively, and it can be
seen from these two figures that (a) both stable and unstable parametric regimes are allowed by the plasma system;
(b) the $k_c$ increases (decreases) with increasing in the value of the negatively (positively) charged
ion mass for a  fixed value of their charge states; (c) the stable parametric regime increases (decreases)
with  positive (negative) ion charge state when their masses remain constant.

The effect of electron number density on the stability condition of IAWs can be understood
by plotting $P/Q$ with $k$ for different values of $\lambda_4$ in Fig. \ref{3Fig:F3}. It is
easy to demonstrate from this figure that the stable (unstable) parametric regime of IAWs
increases (decreases) with the increase in the value of the equilibrium electron number density.
The impact of sub-extensivity and super-extensivity of electrons on the stability condition of IAWs can be seen in
Figs. \ref{3Fig:F4} and \ref{3Fig:F5}, respectively. It is obvious from these two figures that
the sub-extensive property of the electrons allows the IAWs to be stable for large wave number
while the super-extensive property of the electrons allows the IAWs to be stable for small wave number.
\section{Rogue Waves}
\label{3sec:Rogue Waves}
The first-order rogue wave solution of the NLSE can be written as \cite{Akhmediev2009}
\begin{eqnarray}
&&\hspace*{-1.3cm}\Phi(\xi,\tau)=\sqrt{\frac{2P}{Q}}\Big[\frac{4+16i\tau P}{1+4\xi^2+16 \tau^2 P^2}-1\Big] \mbox{exp}(2i\tau P).
\label{3eq:35}\
\end{eqnarray}
It is worthy to mention that the first-order rogue wave solution of the NLSE indicates that a considerable amount of
IAWs energy is condensed into a very small domain in DPP. The effect of the number density and charge state of both
positively and negatively charged ions on the amplitude and width of the IARWs can be observed from Fig. \ref{3Fig:F6}.
It is noted that an increase in the number density of positive (negative) ions tend to enhance (decrease)
both the amplitude and width of the IARWs in the modulationally unstable parametric regime ($P/Q>0$) for a
constant value of positive and negative ions charge state. The physics of this result is that an increase
in the value of positive (negative) ion number density tend to increase (decrease) the nonlinearity as well
as amplitude and width of the IARWs. The nature of IARWs may also be affected by the electron and positron
temperature which can be observed in Fig. \ref{3Fig:F7}. This figure reveals that an increase in the value
of the electron (positron) temperature would make the amplitude and width of the IARWs associated with IAWs
smaller (taller). The physics behind this result is that the nonlinearity of the plasma medium as well as height and width of the IARWs
decreases (increases) with electron (positron) temperature.
\section{Conclusion}
\label{3sec:Conclusion}
We have scrutinized the MI of IAWs and associated IARWs in a four-component DPP having inertial positive and negative
ions and inertialess non-extensive electrons and iso-thermal positrons by deriving a standard NLSE.
It is noted that all of the plasma components in a DPP medium play a vigorous role in the stability criteria of the IAWs.
However, the essence of our findings can be summarized as follows:
\begin{enumerate}
\item{Under consideration of fast and slow mode, both stable and unstable parametric regimes of IAWs can be observed.}
\item{The sub-extensive property of the electrons allows the IAWs to be stable for large wave number
while the super-extensive property of the electrons allows the IAWs to be stable for small wave number.}
\item{An increase in the value of positive (negative) ion number density tend to increase (decrease) the
 nonlinearity as well as amplitude and width of the IARWs.}
\item{The nonlinearity of the plasma medium as well as height and width of the IARWs
decreases (increases) with electron (positron) temperature.}
\end{enumerate}
The implications of our present investigation will be useful in understanding the process
of MI of IAWs and associated IARWs in both laboratory plasma [viz., processing reactors \cite{Gottoscho1986},
semiconductor industry \cite{Shukla1986}, neutral beam sources \cite{Bacal1979}, fullerene ($C^+_{60},C^-_{60}$) \cite{Sabry2008}, and
in intense laser fields \cite{Berezhiani1992}]  and astrophysical
environments [viz., Van Allen radiation belt and near the polar cap of fast rotation neutron
stars \cite{Lightman1982}, solar atmosphere \cite{Tandberg1988}, D-region ($H^+, O_2^-$) and
F-region ($H^+, H^-$) of the earths's ionosphere \cite{Elwakil2010},
upper region of Titan's atmosphere \cite{Labany2012}, etc.].
\section*{Acknowledgement}
S. Jahan gratefully acknowledge NST (National Science and Technology) Fellowship for  their financial support.

\end{document}